\documentstyle[fleqn,12pt]{article}
\begin{document}

\renewcommand{\theequation}{\arabic{section}.\arabic{equation}}
\renewcommand{\thesection}{\Roman{section}}
\baselineskip=24pt
\begin{titlepage}
\vspace*{2cm}

\begin{center}
{\huge   \bf          {\em Yangian}, Truncated {\em Yangian}\\
\vspace*{0.25cm}
                      and Quantum Integrable Models}
\end{center}
\vspace*{1cm}

\begin{center}
{\bf                   M. L. Ge\dag,  Y. W. Wang\dag\ and K. Xue\ddag}\\
\vspace*{0.5cm}
{\small {\it          \dag Theoretical Physics Division,
                       Nankai Institute of Mathematics,}}\\
{\small {\it           Tianjin 300071, People's Republic of China}}\\
{\small {\it     \ddag  Physics Department, Northeast Normal University,}}\\
{\small {\it           Changchun 130024, People's Republic of China }}
\end{center}

\vspace*{0.3in}

\begin{abstract}
\baselineskip=24pt
Based on the RTT relation for the given rational $R$-matrix  the {\em Yangian}
and truncated {\em Yangian} are discussed. The former can be used to generate
the long-range interaction models, whereas the latter can be related to the
Goryachev-Chaplygin(GC) gyrostat. The trigonometric extension of the
Goryachev-Chaplygin gyrostat has also been shown.\\

\end{abstract}
\end{titlepage}

\section{Introduction}
\setcounter{equation}{0}

The RTT relation plays the central role in setting up the connection between
the completely quantum integrable systems and the quantum group symmetry
[1-6]. For a given $R$-matrix satisfying the Yang-Baxter equation(YBE)
variety of operator-valued transfer matrices $T(u)$ can be found to
satisfy the RTT relation. On the one hand the $\mbox{tr}\ T(u)=\sum u^{-n}
h^{(n)}$ gives the conserved quantities $h^{(n)}$ including Hamiltonian,
on the other hand the commutations $[T_{ab}(u),T_{cd}(v)]$ obtained through
RTT relation generate the well-known quantum group. Especially, when
$R$-matrix is rational it defines {\em Yangian} \cite{r1} for some Lie
algebras. From this point of view the RTT relation is an unified description
of both symmetry and physical model that are closely related to each other
for a considered quantum integrable system.

A solution of RTT relation~\cite{r7}
\begin{equation}
{\check R(u-v)}(T(u)\otimes T(v))=(T(v)\otimes T(u)){\check R}(u-v) \label{e1}
\end{equation}
depends on the given form of $\check R(u)$, where
\begin{equation}
\check R(u)=R(u)P\label{e2}
\end{equation}
and $P$ stands for the permutation.

The simplest rational solution of YBE is well-known as~\cite{r8}
\begin{equation}
{\check R}(u)=u P+ I \label{e3}
\end{equation}
where $u$ is the spectral parameter and $I$ the identity. The RTT relation
(\ref{e1}) defines the {\em Yangian} associated with $sl(n)$
algebra~\cite{r1} for
\begin{equation}
T(u)=||T_{ab}(u)||_{a,b=1}^{n}. \label{e4}
\end{equation}
Whereas the operators $T_{ab}$ are given by
\begin{equation}
T_{ab}(u) = \sum^{\infty}_{n=0}u^{-n}T_{ab}^{(n)}.\label{e5}
\end{equation}

The standard procedure to deal with the problem is as the following:
\begin{enumerate}
\item For the given (\ref{e3}) find commutation relations for $T_{ab}^{(n)}$
themselves on the basis of (\ref{e1}) and (\ref{e5}).
\item Re-explain such commutation relations in terms of the generators of
Hopf algebra, for example, $I_{\alpha},\ J_{\alpha}\ (\alpha=1,2,3)$ introduced
by Drinfeld~\cite{r1}.
\item Make special realization of the commutation relations, i.e. specify the
particularly physical objects, satisfying these relations.
\item Obtain the Hamiltonian through such a physical realization.
\end{enumerate}

In this paper we would like to carry out the procedures 1-4 to generate
some quantum integrable models including long-range interaction models [9-19]
and Gyrostat model that are extension of the discussed models in
Refs.~\cite{r21,r22}.

This paper is organized as the following. In section II the general
formulation of the commutation relations for $T_{ab}^{(n)}$ will be given,
based on which we shall distinguish the {\em Yangian} case from the Heisenberg
type of algebraic structure. In section III some long-range interaction
models with internal degree of freedom will be turned out to belong to
the RTT scope, in section IV the GC Gyrostat is shown to belongs to the
Heisenberg type of {\em Yangian} with truncation.
Finally we briefly show some more results in this respect.

\section{Commutation Relations of $T^{(n)}_{ab}$ for $sl(2)$}
\setcounter{equation}{0}

For the given $R(u)$ (\ref{e3}) and the expansion (\ref{e5}) the RTT relation
gives
\begin{equation}
[T^{(0)}_{ab},T^{(n)}_{cd}]=0\ \ \ (a,b,c,d=1,2)\label{f1}
\end{equation}
and
\begin{equation}
[T^{(n+1)}_{bc},T^{(m)}_{ad}]-[T^{(n)}_{bc},T^{(m+1)}_{ad}]+
T^{(n)}_{ac}T^{(m)}_{bd}-T^{(m)}_{ac}T^{(n)}_{bd}=0\label{f2}
\end{equation}

{}From (\ref{f2}) it follows that $T^{(0)}_{ab}$ should be $c$-numbered matrix,
thus it can always be chosen to be diagonal
\begin{eqnarray}
T^{(0)}=\left[\begin{array}{cc}\lambda &0\\ 0& \mu \end{array}\right]\label{f3}
\end{eqnarray}
where $\lambda$ and $\mu$ are arbitrary constants.

After calculations the relations (\ref{f2}) and (\ref{f3}) can be reduced
to the following independent set of relations:\\
{\bf (a)} \vspace*{-1.08cm}
\begin{eqnarray}
&&[\lambda T^{(n)}_{22}-\mu T^{(n)}_{11},T^{(1)}_{12}] =
2\lambda\mu T^{(n)}_{12},\nonumber\\
&&{[\lambda T^{(n)}_{22}-\mu T^{(n)}_{11},T^{(1)}_{21}]} =
-2\lambda\mu T^{(n)}_{21},\label{f4}\\
&&{[T^{(n)}_{12},T^{(1)}_{21}]} =\lambda T^{(n)}_{22}-\mu  T^{(n)}_{11},
\ \ [\lambda T^{(n)}_{22}+\mu T^{(n)}_{11},T^{(1)}_{ab}]=0\nonumber
\end{eqnarray}
for $n=1,2$.\\
{\bf (b)} \vspace*{-1.08cm}
\begin{eqnarray}
&&[\lambda T^{(2)}_{22}+\mu T^{(2)}_{11},\lambda T^{(2)}_{22}-
\mu T^{(2)}_{11}]+ 2\lambda\mu(T^{(1)}_{21}T^{(2)}_{12}-
T^{(2)}_{21}T^{(1)}_{12})=0,\nonumber\\
&&{[\lambda T^{(n)}_{22}+\mu T^{(n)}_{11},T^{(2)}_{12}]}+
T^{(1)}_{12}(\lambda T^{(n)}_{22}-\mu T^{(n)}_{11})-
T^{(n)}_{12}(\lambda T^{(1)}_{22}-\mu T^{(1)}_{11})=0,\label{f5} \\
&&{[\lambda T^{(n)}_{22}+\mu T^{(n)}_{11},T^{(2)}_{21}]}+
T^{(n)}_{21}(\lambda T^{(1)}_{22}-\mu T^{(1)}_{11})-
T^{(1)}_{21}(\lambda T^{(n)}_{22}-\mu T^{(n)}_{11})=0.\nonumber
\end{eqnarray}
{\bf (c)} \vspace*{-1.08cm}
\begin{eqnarray}
&&2\lambda \mu T^{(n+1)}_{12}=[\lambda T^{(n)}_{22}-\mu T^{(n)}_{11},
\ T^{(2)}_{12}]+T^{(1)}_{12}(\lambda T^{(n)}_{22}+\mu T^{(n)}_{11})-
T^{(n)}_{12}(\lambda T^{(1)}_{22}+\mu T^{(1)}_{11}),\nonumber \\
&&\lambda T^{(n+1)}_{22}-\mu T^{(n+1)}_{11}=[T^{(n)}_{12},T^{(2)}_{21}]
+T^{(1)}_{22}T^{(n)}_{11}-T^{(n)}_{22}T^{(1)}_{11},\label{f6}\\
&&2\lambda \mu T^{(n+1)}_{21}=[T^{(2)}_{21},\lambda T^{(n)}_{22}-\mu
T^{(n)}_{11}]+T^{(1)}_{21}(\lambda T^{(n)}_{22}+\mu T^{(n)}_{11})-
T^{(n)}_{21}(\lambda T^{(1)}_{22}+\mu T^{(1)}_{11})\nonumber.
\end{eqnarray}
{\bf (d)}\vspace*{-1.08cm}
\begin{eqnarray}
&&[T^{(n)}_{ab},T^{(m)}_{ab}]=0,\nonumber\\
&&[T^{(n)}_{ab},T^{(m)}_{cd}]=[T^{(m)}_{ab},T^{(n)}_{cd}]\label{f7}.
\end{eqnarray}

{}From (\ref{f3}) one can consider only two possibilities
\begin{eqnarray}
(i) &&\lambda=\mu=1,\label{f8}\\
(ii)&&\lambda=1,\ \mu=0 \label{f9}.
\end{eqnarray}
For the case (i) the set of (\ref{f4})-(\ref{f7}) defines the {\em Yangian}
associated with $sl(n)$. The Drinfeld's {\em Yangian} is
defined by~\cite{r1}
\begin{eqnarray}
\lefteqn{[I_{\lambda},I_{\mu}]=C_{\lambda\mu\nu}I_{\nu}, \ \
[I_{\lambda},J_{\mu}]
=C_{\lambda\mu\nu}J_{\nu},\label{f10}}\\
\lefteqn{[J_{\lambda},[J_{\mu},I_{\nu}]]-[I_{\lambda},[J_{\mu},J_{\nu}]]
=h^2 a_{\lambda\mu\nu\alpha\beta\gamma}\{I_{\alpha},
I_{\beta}, I_{\gamma}\}, \label{f11}}\\
\lefteqn{[[J_{\lambda},J_{\mu}],[I_{\tau},J_{\sigma}]]-
[[J_{\tau},J_{\sigma}],[I_{\lambda},J_{\mu}]]}\nonumber\\
\lefteqn{\ \ =h^2(a_{\lambda\mu\nu\alpha\beta\gamma}C_{\tau\sigma\nu}+
a_{\tau\sigma\nu\alpha\beta\gamma}C_{\lambda\mu\nu})
\{I_{\alpha}, I_{\beta}, I_{\gamma}\},\label{f12}}
\end{eqnarray}
where
\begin{eqnarray}
\lefteqn{a_{\lambda\mu\nu\alpha\beta\gamma}=\frac 1{24}C_{\lambda\alpha\tau}
C_{\mu\beta\sigma}C_{\nu\gamma\rho}C_{\tau\sigma\rho}}\nonumber\\
\lefteqn{C_{\lambda\mu\nu} = i\epsilon _{\lambda\mu\nu}}\label{f14}\\
\lefteqn{\{x_1,x_2,x_3\}=\sum_{\pi}x_{\pi (1)}x_{\pi (2)}x_{\pi (3)}.}\nonumber
\end{eqnarray}
For $sl(2)$ both of the sides of (\ref{f11}) vanish and (\ref{f12}) can be
recast to
\begin{eqnarray}
\lefteqn{[J_3,[J_{+},J_{-}]=\frac{h^2}4 I_3(J_-I_+-I_-J_+)\nonumber}\\
\lefteqn{{[J_{\pm},[J_3,J_{\pm}]]}= \frac {h^2}4 I_{\pm}(J_{\pm}I_3-I_{\pm}J_3)
\label{f16}}\\
\lefteqn{2{[J_3,[J_3,J_{\pm}]]+[J_{\pm},[J_{\pm},J_{\mp}]]}=\frac{h^2}2
I_3(J_{\pm}I_3-I_{\pm}J_3)+I_{\pm}(I_-J_+-J_-I_+)\nonumber}
\end{eqnarray}
where $I_{\pm}=I_1\pm iI_2$ and $J_{\pm}=J_1\pm i J_2$, in terms of which
the (\ref{f10}) read
\begin{eqnarray}
[I_3,I_{\pm}]&=&\pm I_{\pm}, \ \ [I_+,I_-]=2 I_3,\nonumber\\
{[J_3,I_{\pm}]}&=&[I_3,J_{\pm}]=\pm J_{\pm} \label{f18}\\
{[I_{\pm},J_{\mp}]}&=&\pm 2 J_3,\ [I_{\pm},J_{\pm}]=[I_3,J_3]=0
\nonumber
\end{eqnarray}
Now we set
\begin{eqnarray}
\lefteqn{T^{(1)}_{12}=\alpha_+I_+, \ \ T^{(1)}_{21}=\alpha_-I_-, \ \
T^{(1)}_{22}-T^{(1)}_{11}=2I_3,} \nonumber\\
\lefteqn{T^{(2)}_{12}=\beta_+J_+, \ \ T^{(2)}_{21}=\beta_-J_-, \ \
T^{(2)}_{22}-T^{(2)}_{11}=2\beta_3 J_3.\label{f19}}
\end{eqnarray}
Substituting (\ref{f19}) into (\ref{f4}) with $\lambda=\mu=1$ (except for
$[T^{(n)}_{22}+T^{(n)}_{11},T^{(1)}_{ab}]=0$) and (\ref{f7}) with $n,m
=1,2$, one finds that they are satisfied if
\begin{equation}
\alpha_+\alpha_-=1,\ \ \beta_+\alpha_--\alpha_+\beta_-=\beta_3.\label{f20}
\end{equation}
By virtue of (\ref{f20}) substituting (\ref{f19}) into (\ref{f5}) and
(\ref{f6}) where $n=1,2$ and (\ref{f7}) with $n, m=2, 3$ we derive
(\ref{f16}) if
\begin{equation}
h^2=4\beta^{-1}_+\beta^{-1}_-
\end{equation}
The higher-ordered term $T^{(n)}_{ab}$ can be generated by the set of
iteration relation (\ref{f6}).

Obviously the transformation
\begin{equation}
J_{\alpha}\longrightarrow J_{\alpha}+\mbox{const.} I_{\alpha}
\end{equation}
still satisfy the {\em Yangian}.

Therefore, when $\lambda=\mu=1$ we derive the {\em Yangian} associated with
$sl(2)$. Without doubt it is a kind of model-independent symmetry. In order
to obtain the real physical model we have to impose the physical restriction
to the $T(u)$ such that the Hamiltonian can be specified. An
interesting example is the long-range interaction models.

We would like to emphasize that if $\lambda=1$ and $\mu=0$ in (\ref{f3})
it will give rise to the quite different algebraic structure. Under such a
circumstance we have
\begin{equation}
[T^{(n)}_{12},T^{(1)}_{22}]=[T^{(n)}_{21},T^{(1)}_{22}]=[T^{(n)}_{11},T^{(1)}_{22}]=0\label{f22}
\end{equation}
and
\begin{eqnarray}
[T^{(1)}_{12},T^{(n)}_{11}] &=& T^{(n)}_{12}\nonumber \\
{[T^{(1)}_{21},T^{(n)}_{11}]} &=& -T^{(n)}_{21}\label{f23}\\
{[T^{(1)}_{12},T^{(n)}_{21}]} &=& T^{(n)}_{22}\nonumber
\end{eqnarray}
where in the left hand sides of (\ref{f23}) the relations with the
interchange between $(n)$ and $(1)$ are also valid. Obviously,
(\ref{f23}) contains a Heisenberg algebra, for example, by setting
$n=1$ and $T^{(1)}_{22}=0$:
\begin{equation}
T^{(1)}_{11} = i p, \ \ T^{(1)}_{12}=
  e^{-q},\ \  T^{(1)}_{21} =e^{+q},\label{f24}
\end{equation}
if $[p,q]=-i$. This is the reason why we call the algebra Heisenberg type.

For $\mu=0$ and $\lambda=1$ (\ref{f2}) is reduced to the following
independent sets of relations:
\begin{eqnarray}
\lefteqn{[T^{(n)}_{22},T^{(2)}_{11}]+T^{(n)}_{21}T^{(1)}_{12}-
T^{(1)}_{21}T^{(n)}_{12}=0,\ \label{f25}}\\
\lefteqn{T^{(n+1)}_{12}=[T^{(2)}_{12},T^{(n)}_{11}]+T^{(n)}_{11}T^{(1)}_{12}
-T^{(1)}_{11}T^{(n)}_{12},\nonumber} \\
\lefteqn{T^{(n+1)}_{21}=[T^{(n)}_{11},T^{(2)}_{21}]+T^{(n)}_{11}T^{(1)}_{21}
-T^{(1)}_{11}T^{(n)}_{21}, \label{f26}}\\
\lefteqn{T^{(n+1)}_{22}=[T^{(2)}_{12},T^{(n)}_{21}]+T^{(n)}_{11}T^{(1)}_{21}
-T^{(1)}_{11}T^{(n)}_{22}\nonumber}
\end{eqnarray}
for $n\geq 2$ and
\begin{equation}
[T^{(n)}_{ab},T^{(m)}_{ab}]=0,\ \ [T^{(n)}_{ab},T^{(m)}_{cd}]=[T^{(m)}_{ab},
T^{(n)}_{cd}]=0\label{f27}
\end{equation}
where $a,b,c,d=1,2$.

In the section IV we shall return to the algebraic structure of
(\ref{f22})-(\ref{f27}).

\section{RTT Relation and Long-Range Interaction Models}
\setcounter{equation}{0}

Let us apply the {\em Yangian} approach to the long-range interaction  models.
Following the notation of Ref.~\cite{r18} (hereafter it is denoted by BGHP)
the solution of Yang-Baxter equation, $R$-matrix, takes the
simplest form as
\begin{equation}
R_{00^{\prime}}(u)=u+\lambda P_{00^{\prime}}\label{r}
\end{equation}
and the RTT relation (\ref{e1}) reads
\begin{equation}
R_{00^{\prime}}(u-v)T^0(u)T^{0^{\prime}}(v)=
T^{0^{\prime}}(v)T^0(u)R_{00^{\prime}}(u-v)\label{rtt}
\end{equation}
where $T^0(u)=T(u)\otimes 1$, $T^{0^{\prime}}=1\otimes T(u)$ and
$P_{00^{\prime}}$ is the permutation operator exchanging the two auxiliary
spaces $0$ and $0^{\prime}$. Make the expansion \cite{r18}
\begin{eqnarray}
T^0(u) &=& I+\sum^p_{a,b=1}X^0_{ba}\sum^{\infty}_{n=0}
           \frac{\lambda T^{(n)}_{ab}}{u^{n+1}}\ ,\label{t}\\
P_{00^{\prime}} &=& \sum^p_{a,b=1}X^0_{ba}X^{0^{\prime}}_{ab}\ ,\label{p}\\
{[X_{ab},X_{cd}]} &=& \delta_{bc}X_{ad}-\delta_{ad}X_{cb}.
\end{eqnarray}
As was pointed above that $\{T^{(n)}_{ab}\}$ generate the {\em Yangian}
associated with $sl(n)$. Substituting eqs.~(\ref{r}),~(\ref{t}) and
(\ref{p}) into eq.~(\ref{rtt}) one finds
\begin{equation}
\hspace*{-0.5cm}\sum_{a,b}\sum_{cd}X^0_{ba}X^{0^{\prime}}_{dc}\sum^{\infty}
 _{n=0} \left\{ u^{-n-1}f^n_1-v^{-n-1}f^n_2 + \sum^{\infty}_{m=0}u^{-n-1}
 v^{-m-1}f^{n,m}_3 \right\} =0
\end{equation}
where
\begin{eqnarray}
f^n_1 &=& \delta_{bc}T^{(n)}_{ad}-\delta_{ad}T^{(n)}_{cb}-
 [T^{(n)}_{ab},T^{(0)}_{cd}]\ , \nonumber\\
f^n_2 &=& \delta_{bc}T^{(n)}_{ad}-\delta_{ad}T^{(n)}_{cb}-
 [T^{(0)}_{ab},T^{(n)}_{cd}]\ , \nonumber\\
f^{n,m}_3 &=& \lambda(T^{(n)}_{ad}T^{(m)}_{cb}-T^{(m)}_{ad}T^{(n)}_{cb})+
[T^{(n+1)}_{ab},T^{(m)}_{cd}]-[T^{(n)}_{ab},T^{(m+1)}_{cd}]\ .\nonumber
\end{eqnarray}
For any auxiliary space $\{X_{ab}\}$ we require $f^n_1=f^n_2=f^{n,m}_3=0$.
Obviously, $f^n_2=0$ can be induced from $f^n_1=0$. So we need only to take
\begin{equation}
f^n_1=f^{n,m}_3=0\label{f}
\end{equation}
into account.

First from $f^{n,0}_3=0$ it follows
\begin{equation}
\delta_{bc}T^{(n+1)}_{ad}-\delta_{ad}T^{(n+1)}_{cb}=\lambda (T^{(0)}_{ad}
 T^{(n)}_{cb}-T^{(n)}_{ad}T^{(0)}_{cb})+[T^{(n)}_{ab}, T^{(1)}_{cd}]
\end{equation}
which can be recast to
\begin{eqnarray}
\lefteqn{T^{(n+1)}_{ad} =
\lambda(T^{(0)}_{ad}T^{(n)}_{cc}-T^{(n)}_{ad}T^{(0)}_{cc})+[T^{(n)}_{ac},
   T^{(1)}_{cd}]\ (a\neq d)\ , \label{t4}}\\
\lefteqn{T^{(n+1)}_{aa}-T^{(n+1)}_{cc} = \lambda(T^{(0)}_{aa}T^{(n)}_{cc}-
  T^{(n)}_{aa}T^{(0)}_{cc}) +[T^{(n)}_{ac}, T^{(1)}_{ca}]\ , \label{t2}}
\end{eqnarray}
where no summation for the repeating indices is taken. Eqs.~(\ref{t4}) and
(\ref{t2}) imply that $T_{ab}^n$ can be determined by iteration for
given $T^{(0)}_{ab}$ and $T^{(1)}_{ab}$.

Now let us set
\begin{eqnarray}
T^{(0)}_{ab} &=& \sum^N_{i=1}I^i_{ab}\ , \label{ti} \\
T^{(1)}_{ab} &=& \sum^N_{i=1}I^i_{ab}D_i \label{d}
\end{eqnarray}
and
\begin{equation}
[I^i_{ab}, I^j_{cd}]= \delta_{ij}(\delta_{bc}I^i_{ad}-\delta_{ad}I^i_{cb})
\label{ij}
\end{equation}
where $D_i$ are operators to be determined.  $\{I^i_{ab}\}$ can be
understood as internal degree of freedoms, such as spins. Substituting
eqs.~(\ref{ti})--(\ref{ij}) into $f^1_1$ we obtain
\begin{equation}
\sum_i\sum_j I^i_{ab}[D_i, I^j_{cd}]=0\ .
\end{equation}
Further we assume
\begin{equation}
\sum_i I^i_{ab}[D_i, I^j_{cd}]=0,\ \ \mbox{for\ any}\ j \label{di}
\end{equation}
with which the $T^{(2)}_{ab}$ should satisfy
\begin{eqnarray}
\delta_{bc}T^{(2)}_{ad}-\delta_{ad}T^{(2)}_{cb} &=& \sum_{i\neq
j}I^i_{ab}I^j_{cd}
    \left\{ \lambda \sum_{k,l}I^i_{kl}I^j_{lk} (D_j-D_i)+[D_i,D_j] \right\}
    \nonumber\\
&&+\sum_i(\delta_{bc}I^i_{ad}D^2_i-\delta_{ad}I^i_{cb}D^2_i)\label{t3}\ .
\end{eqnarray}
A sufficient solution of eq.~(\ref{t3}) is
\begin{equation}
T^{(2)}_{ab}=\sum_i I^i_{ab}D^2_i
\end{equation}
with
\begin{equation}
[D_i, D_j]=\lambda \sum_{a,b}I^j_{ab} I^i_{ba}(D_i-D_j)\label{did}\ .
\end{equation}
Thus eq.~(\ref{d}) generates long-range interaction through the
eq.~(\ref{di}) and (\ref{did}).
However so far there is not simple relationship between $D_i$ and $I^j_{ab}$
which should satisfy eq.~(\ref{di}). It is very difficult to determine
the general relationship. Fortunately, BGHP have set up the link
with the help of projection. Let the permutation groups $\Sigma_1$,
$\Sigma_2$ and $\Sigma_3$ be generated by $K_{ij}$, $P_{ij}$ and
the product $P_{ij}K_{ij}$ respectively, where $K_{ij}$ exchange
the positions of particles and $P_{ij}$ exchange the spins at position
$i$ and $j$. The projection $\rho$ was defined as
\begin{equation}
\rho (ab)=a \ \ \mbox{for}\ \forall a\in \Sigma_2, b\in \Sigma_1\ ,
\end{equation}
i.e. the wave function considered is symmetric. Let $I^i_{ab}$ be the
fundamental representations, then
\begin{equation}
P_{ij}=\sum_{a,b}I^i_{ab} I^j_{ba}\ .
\end{equation}
Suppose that there exists
\begin{equation}
D_i=\rho ({\hat D}_i), \ \ D_i \in \Sigma_2,\ {\hat D}_i\in \Sigma_1
\label{e21}
\end{equation}
and the ${\hat D}_i$ is particle-like operators, i.e.
\begin{equation}
K_{ij}{\hat D}_i={\hat D}_jK_{ij}, \ \ K_{ij}{\hat D}_l={\hat D}_lK_{ij}
\ \  (l\neq i,j)\ .
\end{equation}
Define
\begin{equation}
T^{(m)}_{ab}=\sum_i I^i_{ab} \rho ({\hat D}^m_i) \ \ (m\geq 0)\ ,\label{tt}
\end{equation}
then\\
\begin{equation}\hspace*{-1cm}\mbox{(a)}\ \ \
[{\hat D}_j, {\hat D}_i] = \lambda \rho^{-1}(P_{ij}(D_j-D_i))
        = \lambda ({\hat D}_j-{\hat D}_i)K_{ij}\ . \label{dd}
\end{equation}
(b) $T^{(m)}_{ab}$ satisfy eq.~(\ref{f}), i.e., RTT relation eq.~(\ref{rtt}).

Actually $f^n_1=0$ is easy to be checked. By using
\[
[{\hat D}^n_i, {\hat D}^m_j] = \sum^{n-1}_{k=0}{\hat D}^k_i[{\hat D}_i,
       {\hat D}^m_j]{\hat D}^{n-k-1}_j = \lambda \sum^{n-1}_{k=0}
       {\hat D}^k_i({\hat D}^m_i-{\hat D}^m_j){\hat D}^{n-k-1}_jK_{ij}\ ,
\]
we have $f^{n,m}_3=0$.

The projection procedure is very important for it enables us to prove that
eq.~(\ref{f}) is satisfied by virtue of eq.~(\ref{e21}).

With the expansion eq.~(\ref{t}) and the projected long-range
expansion eq.~(\ref{tt}), the hamiltonian associated to $T(u)$ is
obtained by the expansion of the deformed determinant \cite{r1,r18}:
\begin{equation}
det_qT(u)=\sum_{\sigma} \epsilon(\sigma)T_{1\sigma_{1}}(u-(p-1)\lambda)
T_{2\sigma_{2}}(u-(p-2)\lambda)\cdots T_{p\sigma_{p}}(u)\ .
\end{equation}
A calculation gives
\begin{eqnarray}
det_qT(u) &=& 1+\frac {\lambda}uM+\frac{\lambda}{u^2}\left[
  \rho(\sum_i{\hat D}_i-\frac{\lambda}2 \sum_{j\neq i}K_{ij})+
  \frac{\lambda}2 M(M-1)\right]\nonumber\\
&& +\frac{\lambda}{u^3}\rho\left\{ (\sum_i{\hat D}_i-\frac{\lambda}2\sum_
  {j\neq i}K_{ij})^2+\frac{\lambda^2}{12}\sum_{i\neq j\neq k\neq i}K_{ij}K_{jk}
  \right. \nonumber\\
&& +\lambda (M-1)\sum_i ({\hat D}_i-\frac{\lambda}2\sum_{j\neq i}K_{ij})
  \nonumber\\
&& +\left.\frac{\lambda^2}6M(M-1)(M-2)+\frac{\lambda^2}4M(M-1)\right\}
   +\cdots\ .\label{det}
\end{eqnarray}
One takes the Hamiltonian as
\begin{equation}
H=\frac 12 \rho \left\{(\sum_i{\hat D}_i-\frac{\lambda}2 \sum_{i\neq j}K_
  {ij})^2+\frac{\lambda^2}{12}\sum_{i\neq j\neq k\neq i}K_{ij}K_{jk}\right\}
  \label{g27}\ .
\end{equation}
Therefore we define the Hamiltonian which have the {\em Yangian}
symmetry given by eqs.~(\ref{ij}), (\ref{did}) and (\ref{tt}). The
Hamiltonian~(\ref{g27}) is concided with the results of Refs.~\cite{r13,r14}.
In comparison to the known models we list the expressions for ${\hat D}_i$
satisfying eq.~(\ref{dd})

\noindent{\bf (1)} \hspace*{0.5cm} $ {\hat D}_i = p_i+\frac{\lambda}2
 \sum_{i\neq j}[\mbox{sgn}(x_i-x_j)+1]K_{ij}, \ \ \lambda=2il$,
\begin{equation}
H= \frac 12 \sum_ip^2_i +\frac 12 \sum_{i\neq j}l(l-P_{ij})
      \delta(x_i-x_j)\ .\label{h0}
\end{equation}

\noindent{\bf (2)} \hspace*{0.5cm} $ {\hat D}_i = p_i+\sum_{i\neq j}l
 [i\cot a(x_i-x_j)+1]K_{ij},\  \lambda=2l$,
\begin{equation}
H = \frac 12 \sum_ip^2_i +\frac 12 \sum_{i\neq j}\frac{l(l-aP_{ij})}
{\sin^2a(x_i-x_j)}\ .\label{h02}
\end{equation}
This is the famous Calagero-Sutherland model with an internal degree
of freedom.

\noindent{\bf (3)}\hspace*{0.5cm}${\hat D}_i = p_i+il\sum_{i\neq j}
[\coth a(x_i-x_j)P^+_{ij}+\tanh a(x_i-x_j)P^-_{ij}+1]K_{ij}, \ \lambda=2il\ , $
\begin{equation}
H = \frac 12 \sum_ip^2_i +\frac 12 \sum_{i\neq j}l(l-aP_{ij})
 \left(\frac{P^+_{ij}}{\sinh^2a(x_i-x_j)}-\frac{P^-_{ij}}{\cosh^2a(x_i-x_j)}
 \right)\label{h3}
\end{equation}
where $P^{\pm}_{ij}=\frac{1\pm \sigma_i\sigma_j}2,\ \sigma$ is quantum number
with $\sigma^2=1$.
At the case of $P_{ij}=1$, eq.~(\ref{h3}) was first introduced by
Calogero etc.~\cite{rc} and studied recently by Sutherland and R\"omer
in Ref.~\cite{r19}.

An alternative description of transfer matrix was given by BGHP.
Define
\begin{equation}
{\bar D}_i={\hat D}_i-\lambda \sum_{i<j}K_{ij}\ ,
\end{equation}
then
\begin{eqnarray}
[{\bar D}_i, {\bar D}_j] &=&0\ , \label{d1}\\
\left[ K_{ij},{\bar D}_k \right] &=&0\ \ (k\neq i,j)\ ,\label{d2}\\
K_{ij}{\bar D}_i-{\bar D}_jK_{ij} &=& \lambda\ .
\end{eqnarray}
It was proved that
\begin{equation}
{\bar T}_i(u)=1+\lambda \frac{P_{0i}}{u-{\bar D}_i},\ {\bar T}(u)
=\prod_i{\bar T}_i(u)\ \ \mbox{and}\ \ \rho({\bar T}(u))\label{t1}
\end{equation}
all satisfy the RTT relation.

The deformed determinant of ${\bar T}(u)$ was defined by
\begin{equation}
det_q{\bar T}(u)=\frac{\Delta_M(u+\lambda)}{\Delta_M(u)}\ ,\ \
\Delta_M(u)=\prod^M_{i=1}(u-{\bar D}_i)\ .\label{g35}
\end{equation}
It was proved that
\begin{equation}
\rho(det_q{\bar T}(u))=det_q(T(u))\ .
\end{equation}

We can give another kind ${\hat D}_i$ satisfies eqs.(\ref{d1}),
and (\ref{d2}) etc.
\begin{equation}
[\hat {D}_i, \hat {D}_j] = 2\beta P^+_{ij}
(\hat {D}_i-\hat {D}_j)K_{ij}\label{g36}\
\end{equation}
and
\begin{equation}
{\bar D}_i={\hat D}_i-\beta\sum_{j<i}P^+_{ij}K_{ij} \label{g37}
\end{equation}
The Hamiltonian can be obtained from eq.(\ref{g35}) or from the following
equation
\begin{equation}
H=\frac 12 \rho\left\{ \sum_i {\bar D}^2_i+\frac {\beta^2}6 \sum_{i\neq j
\neq k\neq i} P^+_{ijk}K_{ij}K_{jk}\right\}\ .\label{g38}
\end{equation}
where $P^+_{ijk}=P^+_{ij}P^+_{jk}$. Now we have two sufficient solutions of
${\hat D}_{ij}$:

\noindent{\bf (1)} \hspace*{0.5cm} $ {\hat D}_i = p_i+\sum_{i\neq j}l
 [i\cot a(x_i-x_j)+1]K_{ij}P^+_{ij},\  \lambda=2l$,
\begin{equation}
H = \frac 12 \sum_ip^2_i +\frac 12 \sum_{i\neq j}\frac{l(l-aP_{ij})}
{\sin^2a(x_i-x_j)}P^+_{ij}\ .\label{g41}
\end{equation}
Eq.~(\ref{g41}) is the generalization of the spin chain model~(\ref{h02})
considered by BGHP \cite{r18}.

\noindent{\bf (2)} \hspace*{0.5cm} $ {\hat D}_i = p_i+\frac{\lambda}2
 \sum_{i\neq j}[\mbox{sgn}(x_i-x_j)+1]K_{ij}P^+_{ij}, \ \ \lambda=2il$,
\begin{equation}
H= \frac 12 \sum_ip^2_i +\frac 12 \sum_{i\neq j}l(l-P_{ij})
      \delta(x_i-x_j)P^+_{ij}\ .\label{g40}
\end{equation}
On condition that $P_{ij}=\pm 1$, eq.~(\ref{g40}) was first pointed out
by Yan \cite{r20} through Bethe Ansatz, he also found the Y-operator
defined by Yang for his model
\begin{equation}
Y^{\alpha \beta}_{ij}=\frac{1}{ik_{ij}(ik_{ij}-c)}[ik_{ij}- cP^-_{ij}]
[-ik_{ij}P^{\alpha \beta}+ c P^+_{ij}]
\end{equation}
where $c=l(l\pm 1)$, $P^{\alpha\beta}$ is the permutation,
and $Y$ satisfies~\cite{r8}
\begin{equation}
Y^{\alpha \beta}_{jk}Y^{\beta \gamma}_{ik}Y^{\alpha \beta}_{ij}=
Y^{\beta \gamma}_{ij}Y^{\alpha \beta}_{ik}Y^{\beta \gamma}_{jk}
\end{equation}

In conclusion of this section we have shown the consistence between
{\em Yangian} symmetry and the integrability of Polychronakos for
long-range interaction models and given the interpretation of model~(\ref{h3})
and Yan model~(\ref{g40}) from the point of view of YB system.

\section{Truncated {\em Yangian} and GC Gyrostat}
\setcounter{equation}{0}

In this section we shall show that by making the truncation of the expansion
(\ref{e5}) at $n=4$ (i.e. $T^{(4)}_{ab}=0$) and set $\mu=0$ in (\ref{f3})
the set of commutation relation (\ref{f22})-(\ref{f27}) is equivalent to the
Goryachev-Chaplygin(GC) gyrostat.

The Goryachev-Chaplygin(GC) top is completely integrable in sense of
Liouville as shown in the literature~\cite{r21,r22}. In the quantum case the GC
top is extended to the quantum GC gyrostat whose Hamiltonian takes the
form~\cite{r21,r22}:
\begin{equation}
H=\frac 12 (J^2+3J^2_3)-bx_1+pJ_3\label{h1}
\end{equation}
where $J^2=J^2_1+J^2_2+J^2_3$ and $b$ are parameters. The quantities
appearing in eq.~(\ref{h1}) satisfy the following commutation relations:
\begin{eqnarray}
[J_i, J_j]   &=& -i\epsilon_{ijk}J_k\ ,\nonumber\\
{[J_i,x_j]}  &=& -i\epsilon_{ijk}x_k\ ,\nonumber\\
{[x_i,x_j]}  &=& 0\ ,\label{h2}\\
{[p,q]}      &=& -i\ ,\nonumber\\
{[p,J_i]}     &=& [p,x_i]=[q,x_i]=[q,J_i]=0\nonumber
\end{eqnarray}
where $i,\ j,\ k$ take over $1,\ 2$ and $3$. The momentum $p$ and coordinate
$q$ of mass center commute with all the locally dynamic variables.
The Hamiltonian eq.~(\ref{h1}) commutes with another integral of motion:
\begin{equation}
G=(2J_3+p)(J^2-J^2_3)+2b[x_3,J_1]_+
\end{equation}
where $[A,B]_+\equiv AB+BA$.
In addition to the commutation relations eq.~(\ref{h2}) there are two
constraints \cite{r22}:
\begin{eqnarray}
\sum^3_{i=1}J_ix_i &=& 0 \label{h4},\\
\sum^3_ix^2_i &=& 1  \label{h5}.
\end{eqnarray}
Eq.~(\ref{h4}) plays important roles in proving the quantum integrability of
the model.

Sklyanin has pointed out that the GC gyrostat can be systematically
investigated in the framework of RTT relation \cite{r22}.
It is interesting to re-derive the GC Gyrostat on
the basis of the detail analysis of {\em Yangian} and explore some points
not touched in~\cite{r22}, for example, the derivation of more general
solution of the transfer matrix $T(u)$ in terms of $R$-matrix formalism.

In this section we shall mainly focus on the two points. The first is to
show that the GC gyrostat represents a new type of {\em Yangian} which
can be called ``truncated {\em Yangian}''. The second goal is to look
for more general form of the transfer matrix $T(u)$ on the basis of
$R$-matrix formalism.

For solving (\ref{f22})-(\ref{f27}) with $T^{(n)}_{ab}=0$ for $n\geq 4$ we
first put~\cite{r22}
\begin{equation}
T^{(1)}_{11} = \alpha p, \ \ T^{(1)}_{22} =0,\ \ T^{(1)}_{12}=
  \beta e^{\tau q}x_+ ,\ \  T^{(1)}_{21} =\gamma e^{-\tau q},\label{h17}
\end{equation}
and
\begin{eqnarray}
T^{(2)}_{11} &=& f_1J^2+f_2J^2_3+f_3pJ_3+f_4x_-+f_5,\nonumber\\
T^{(3)}_{11} &=& (p+g_3J_3)(g_1J^2+g_2J^2_3+g_5)+g_4[J_-,x_3]_+\label{h18}
\end{eqnarray}
where $f_1,\ldots,f_5,\ g_1,\ldots,g_4,\ \alpha,\ \beta,\ \gamma$ and $\tau$
all parameters to be determined and $J_{\pm}=J_1\pm iJ_2$, $x_\pm=x_1
\pm ix_2$ obey the commutation relations shown by eq.~(\ref{h2}).
Obviously not all the parameters are independent.
$\lambda$ can be normalized to be one. Substituting
eqs.~(\ref{h17}) and (\ref{h18}) together with eq.~(\ref{h4}) into
eq.~(\ref{f22})-(\ref{f27}) after lengthy calculations by hand we find
\begin{eqnarray*}
\lefteqn{ \tau=-i\alpha^{-1}\lambda,\ f_1=-\frac 14\lambda,
   \ f_2=-\frac 34\lambda,\ f_3=\alpha,\ f_5=-\frac 1{16}\lambda,} \\
\lefteqn{ g_1=-g_2=-\frac 14 \alpha,\ g_3=-\lambda\alpha^{-1},
   \ g_4=\frac 14f_4, \ g_5=-\frac 1{16}\alpha.}
\end{eqnarray*}
Denoting $f_4=f$ the solution $T^{(n)}_{ab}$ reads:
\begin{eqnarray}
T^{(2)}_{11} &=& -\frac 14 \lambda (J^2+3J^2_3+\frac 14)+\alpha p J_3+f x_-,
\ \ T^{(2)}_{22}= \lambda^{-1}\gamma \beta x_+,\nonumber\\
T^{(2)}_{12} &=& - \lambda^{-1}\beta e^{\tau q}\left( -\frac{\lambda}4
  [J_+, x_3]_+ +x_+(\lambda J_3-\alpha p)\right),\ \ T^{(2)}_{21} =
  \gamma e^{-\tau q}J_3,\nonumber\\
T^{(3)}_{11} &=& -\frac 14 \alpha (p-\lambda \alpha^{-1}J_3)(J^2-J^2_3+
  \frac 14) +\frac 14 f[J_-, x_3]_+,\nonumber\\
T^{(3)}_{22} &=& \frac 14 \lambda^{-1}\beta \gamma [J_+, x_3]_+,\label{h25}\\
T^{(3)}_{12} &=& -\lambda^2\beta e^{\tau q}\left\{ fx^2_3-\frac 14 \alpha
[J_+, x_3]_+(p-\lambda \alpha^{-1}J_3)\right\},\nonumber\\
T^{(3)}_{21} &=& -\frac 14\gamma e^{-\tau q}(J^2-J^2_3+\frac 14)\nonumber.
\end{eqnarray}
Therefore the truncated {\em Yangian} can be viewed as a mapping of the algebra
given by eq.~(\ref{h2}).

Let us calculate the Casimirs through
\begin{eqnarray}
\mbox{det}_qT(u)&=&T_{11}(u)T_{22}(u-1)-T_{12}(u)T_{21}(u-1)\nonumber\\
&=&\sum^{\infty}_{n,l=0,m=1}u^{-n-m-l}\frac{(l+m-1)!}{(n-1)!l!}
 (T^{(n)}_{11}T^{(m)}_{22}-T^{(n)}_{12}T^{(m)}_{21})\nonumber\\
&\equiv&\sum^{\infty}_{i=0}u^{-i}C_i
\end{eqnarray}
it gives
\begin{eqnarray}
C_0 &=& 0,\  C_1=\lambda T^{(1)}_{22}, \nonumber\\
C_j &=& \lambda T^{(j)}_{22}+\lambda \sum_{\scriptsize
\begin{array}{c}m+l=j\\m,l\neq 0
 \end{array}} \frac{(l+m-1)!}{(m-1)!l!}T^{(m)}_{22}\nonumber\\
 &&+\sum_{\scriptsize \begin{array}{c}m+n+l=j\\m,n\neq 0 \end{array}}
\frac{(l+m-1)!}
 {(m-1)!l!}\left( T^{(n)}_{11}T^{(m)}_{22}-T^{(n)}_{12}T^{(m)}_{21}\right)
 \ (j\geq 1)\label{h19}.
\end{eqnarray}
Substituting the derived $T(u)$ into eq.~(\ref{h19}) we obtain
\begin{equation}
\mbox{det}_qT(u)=u^{-3}(u-1)^{-3}\lambda^{-1}f(u^2-u+\frac 3{16})
  (x_+x_-+x^2_3).
\end{equation}
So only $\sum^3_{i=1}x^2_i$ is a Casimir operator, i.e. $C_m=0$ for $m<4$,
$C_4=\lambda^{-1}f\sum^3_{i=1}x^2_i$ that commute with $J_i$. The
constraint eq.~(\ref{h5}) does not play role in solving the RTT relation.
By choosing a proper representation eq.~(\ref{h5}) can be taken that
is automatically satisfied by virtue of $[\mbox{det}_q T(u),T_{ab}(v)]=0$.
Notice that $\mbox{det}_qT(u)$ has zero's at $u=\frac 14$ and
$u=\frac 34$ where the inverse of $T(u)$ can not be defined.

It is worth noting that the commutation relations shown by eq.~(\ref{h2})
are invariant subject to a transformation:
\begin{equation}
J^{\prime}_a=\sum_b A_{ab}J_b,\ x^{\prime}_a=\sum_b A_{ab}x_b,
\ \ (a,b=1,2)\label{h36}
\end{equation}
where
\begin{eqnarray}
\begin{array}{ll} A_{11}=\epsilon A_{22}, &A_{12}=-\epsilon A_{21},\\
A^2_{11}+A^2_{21}=1, & \epsilon=\pm 1=\mbox{det}A. \end{array}
\end{eqnarray}
This transformation is useful to transform the Hamiltonian in preserving the
RTT relation.

Let us turn to the conserved quantities for GC gyrostat.

By taking the trace of the transfer matrix one obtains
\begin{eqnarray}
\hspace{-1.2cm}
\mbox{tr}T(u)&=&\lambda+u^{-1}\alpha p+u^{-2}\left\{-\frac 14 \lambda
   (J^2+3J^2_3+\frac 14)+\alpha p J_3+fx_-+\lambda^{-1}\beta\gamma x_+
   \right\}+\nonumber\\
&&\frac {u^{-3}}4\left\{-(\alpha p-\lambda J_3)(J^2-J^2_3+
   \frac 14)+ f[J_-,x_3]_++\lambda^{-1}\beta\gamma
   [J_+,x_3]_+\right\}.
\end{eqnarray}
Hence we have the Hamiltonian $H_p$ and another conserved quantities $G_p$:
\begin{eqnarray}
H_p &=& \frac 12\left\{\frac 14\lambda (J^2+3J^2_3)-fx_--\lambda^{-1}\beta
  \gamma x_+-\alpha p J_3 \right\}+\frac 1{16}\lambda \label{h29},\\
G_p &=& \frac 14 \left\{(\alpha p-\lambda J_3)(J^2-J^2_3+\frac 14)+
   f[J_-,x_3]_++\lambda^{-1}\beta\gamma[J_+,x_3]_+\right\}\label{h30}.
\end{eqnarray}
If one requires
\begin{equation}
4\lambda^{-1}\beta\gamma f=b^2,\label{h31}
\end{equation}
then there is rotational invariance about the $x_3$ axis. By virtue of
eqs.~(\ref{h25}) and (\ref{h36}) we find that eqs.~(\ref{h29}) and (\ref{h30})
are reduced to
\begin{eqnarray}
H_p &=& \frac 12\left\{\frac 14\lambda ({J^{\prime}}^2+3{J^{\prime}}^2_3)
   -b{x^{\prime}}_1-\alpha p\epsilon {J^{\prime}}_3\right\}+
   \mbox{const.}\\
G_p &=& -\frac 14 \left\{(\alpha p-\epsilon\lambda {J^{\prime}}_3)
   ({J^{\prime}}^2-{J^{\prime}}^2_3+\frac 14)+b[x_3,{J^{\prime}}_1]_+
   \right\}
\end{eqnarray}
which are exactly those given by Sklyanin~\cite{r22}. Obviously,
the parameters $\lambda$ and $\tau$ are trivial in the solution of
$T(u)$ and $\beta$, $\gamma$ can be viewed as the consequence of
$T(u)$ subject to a similar transformation. Therefore only parameters
$f$ and $\alpha$ are essential in determining the form of $T(u)$
though $f$ does not appear in the Hamiltonian. The simplest realization of
quantum GC top is a quantum mechanical system on sphere.

Taking the constraints eqs.~(\ref{h4}) and (\ref{h5}) into account the
simplest realization of $J_i$ and $x_i$ can be made through
\begin{eqnarray}
x_i &=& n_i(t),\\
J_i &=& -i\epsilon_{ijk}n_j(t)\dot n_k(t)\ (i,j,k=1,2,3)
\end{eqnarray}

that give $J^2=-\dot{\vec n}\cdot\dot{\vec n}$.

In conclusion we have shown that the truncated {\em Yangian} gives rise to
quantum integrable system with finite number of conserved quantities and
make the system work in higher than (1+1) dimensions.

The trigonometric extension of the truncated {\em Yangian} will be the
``truncated'' affine quantum algebra which is the generalization of
Drinfeld's statement discussed in Ref.~\cite{r3}. It gives
$q$-deformed GC gyrostat. In the following section, we will show it.

\section{Further Development}
\setcounter{equation}{0}

Along the same lines discussed in the section III and IV we are able to
develop the idea: for a given $R$-matrix find the corresponding
$T(u)$-matrix.

\noindent{\bf (1)} The first extension should be made for the $R$-matrix
given by the standard 6-vertex form whose rational limit leads to (\ref{e3}).
The
trigonometric $R$-matrix takes the form:
\begin{eqnarray}
{\check R}_T(x)=\left[ \begin{array}{cccc}a(x)& & & \\ &w&b(x)&  \\
 &b(x)&w& \\ & & &a(x)\end{array}\right] \label{g31}
\end{eqnarray}
\begin{equation}
{\check R}_T(xy^{-1})(T(x)\otimes T(y))=
(T(y)\otimes T(x)){\check R}_T(xy^{-1}) \label{g32}
\end{equation}
where $a(x)=qx-q^{-1}x^{-1}$, $b(x)=x-x^{-1}$, $w=q-q^{-1}$, $x=e^u$
and $q=e^\gamma$. Following Drinfeld~\cite{r3} the transfer matrix $T(u)$
obeying (\ref{g32}) for ${\check R}={\check R}_T$
should have the expansion form
\begin{equation}
T(x)=\sum^{+\infty}_{n=-\infty}x^nT^{(n)} \label{g33}
\end{equation}
We want to find the truncated (affine) solution $T^{(u)}\ (|n|\leq 3)$,
i.e., which possesses the form $T(x)=\sum^{3}_{n=-3}x^nT^{(n)}$ which
takes the solution (\ref{h17})-(\ref{h25}) as a rational limit. We shall
see that this extension will give rise to noncommutative geometry rather
than those discussed in Ref.~\cite{r22}.

After tedious calculations we find that the algebra for truncated
$T^{(n)}_{ab}\ (|n|\leq 3)$ in (\ref{g33}) is equivalent to the algebra
${\cal A}^q$ formed by~\cite{r23} $[P,Q]=-i$, ($P,\ Q$ commute with ${\hat J}
_{\pm},\ {\hat x}_{\pm},\ {\hat J}_3$ and ${\hat x}_3$) and
\begin{eqnarray}
[{\hat J}_{\pm},{\hat x}_{\pm}] &=& [{\hat J}_3,{\hat x}_3] =0,\nonumber\\
{[{\hat J}_{\pm}, {\hat J}_3]} &=& \pm {\hat J}_3,\ \
[{\hat J}_+, {\hat J}_-] =-g[{\hat J}_3]_q,\nonumber\\
{[{\hat x}_{\pm},{\hat J}_3]} &=& \pm {\hat x}_{\pm},\nonumber\\
q^{\delta_{\mp}}{\hat J}_{\pm}{\hat x}_3 &=& {\hat x}_3{\hat J}_{\pm}
\pm \tau_{\pm}K^{\pm\delta_{\pm}}{\hat x}_{\pm}\label{g18},\\
{[{\hat x}_{+},{\hat x}_{-}]} &=& 0,\ \ \
{\hat x}_{\pm}{\hat x}_{3}=q^{\delta_{\pm}}{\hat x}_3{\hat x}_{\pm},\nonumber\\
q^{-1}{\hat J}_{\pm}{\hat x}_{\mp} &=& {\hat x}_{\mp}{\hat J}_{\pm}
\mp \tau^{-1}_{\mp}gK^{\mp\delta_{\pm}}{\hat x}_{3}\nonumber
\end{eqnarray}
with $\delta_++\delta_-=\pm 1$, and the constraint condition
\begin{equation}
g[{\hat J}_3]_q{\hat x}_3+\tau_-K^{-\delta_-}{\hat J}_+{\hat x}_-+\tau_+
K^{\delta_+}{\hat J}_-{\hat x}_+=0\label{g19}
\end{equation}
when $q\rightarrow 1$, ${\cal A}^q$ returns to (\ref{h2}), namely (\ref{g18})
is nothing but the trigonometric extension of (\ref{h2}).

The expression of $T^{(n)}_{ab}(n\leq 3)$ in terms of $\{ {\hat J}_{\pm},
\ {\hat x}_{\pm},\ {\hat J}_3,\ {\hat x}_3\}$ is explicitly given in
Ref.~\cite{r23}. Also the Hamiltonian $H$ and another conserved quantity
$G$ have the forms:

\begin{eqnarray}
H &=& 2i\lambda_3\{\alpha \alpha^{-1}_2(q-1)wg^{-1}\cos[\eta(P+\xi
{\hat J}_3)]{\hat J}_+{\hat J}_-\nonumber\\
&&+2\sin(\xi\eta{\hat J}_3)(\sin[\eta(P+\xi {\hat J}_3]-\alpha^{-1}\alpha_2
q^{\frac 12}\sin[\eta (P+\xi {\hat J}_3+\frac 12)])\nonumber\\
&&+(\alpha\alpha^{-1}_2(q+1)+2)\cos (\xi P)\}+D_+  \label{i8},
\end{eqnarray}
\begin{eqnarray}
G &=&-2i\lambda_3\left\{\alpha \alpha^{-1}_2(q-1)wg^{-1}{\hat J}_+
{\hat J}_-+2(\alpha\alpha^{-1}_2q^{\frac 12}\cos[\xi\eta ({\hat J}_3+\frac
12)])
\right.\nonumber\\
&&+\left.\left.\cos (\xi\eta {\hat J}_3)\right\}\sin[\eta(P+\xi
{\hat J}_3]+D_- \right) \label{i9}
\end{eqnarray}
where $q=e^{i\xi\eta}$ and $D_{\pm}$ are given by
\begin{eqnarray}
D_{\pm} &=& \epsilon_{\pm}\left\{\lambda^{-1}_3\alpha_2\beta^{(1)}_+(\pm{\hat
J}_{\pm}{\hat x}_3K^{-1}+(1-q^{-1})^{-1}q^{-2+\delta_+}\tau_+K^{\delta_+-1}
(\alpha^{-1}\alpha_2q^{-1}\mp 1){\hat x}_+)\right.\nonumber\\
&&+\left.\lambda^{(1)}_+({\hat J}_-{\hat
x}_3+(1-q^{-1})^{-1}q^{-\delta_+}\tau_-
K^{1+\delta_+}(1\mp q^{-1}\alpha^{-1}\alpha_2){\hat x}_-)\right\}\label{i10}
\end{eqnarray}
for $\delta_++\delta_-=1$, and
\begin{eqnarray}
D_{\pm} &=& \epsilon_{\pm}\left\{\lambda^{-1}_3\alpha_2\beta^{(1)}_+({\hat
J}_{\pm}{\hat x}_3K+(1-q)^{-1}q^{2+\delta_+}\tau_+K^{\delta_++1}(\alpha
\alpha^{-1}_2q\mp 1){\hat x}_+)\right.\nonumber\\
&&+\left.\lambda^{(1)}_-(\mp{\hat J}_+{\hat x}_3+(1-q)^{-1}q^{-\delta_+}
\tau_-K^{1+\delta_+}(\alpha\alpha^{-1}_2q\mp 1){\hat x}_-)\right\}\label{i11}
\end{eqnarray}
for $\delta_++\delta_-=-1$, with $\epsilon_+=i,\ \epsilon_-=1$.

It is noted that
\begin{equation}
\alpha=\pm q^{-\frac 12}\alpha_2 \label{i12}
\end{equation}
in (\ref{i10}) and (\ref{i11}).

We would like to remark that for the given standard 6-vertex $R$-matrix
(\ref{g31}) we find a set of solution for truncated RTT relation~\cite{r23}.
The solution can be realized through the algebra (\ref{g18}) and
corresponding conserved quantities (\ref{i8}) and (\ref{i9}).
Eq.~(\ref{g18}) naturally yields the non-commutative geometry. Since the
considered system is axially symmetric so that  the coordinates on $(x_1,
\ x_2)$ plane are commute with each other, whereas the third coordinates
$x_3$ does not commute with them.

\noindent{\bf (2)} For the {\em Yangian} case ($\lambda=\mu=1$ in (\ref{f3}))
more
possibility of realizations of the transfer matrix $T^{(n)}_{ab}$ in
(\ref{e5}) can be found.

The interesting examples are those expressed in terms of the second
quantization forms. They allow to obtain some physical Hamiltonian
from Yang-Baxter systems expressed by annihilation and creation operators.
For instance we may ask whether the Hamiltonian of 1-dim. Hubbard model
can be obtained from the RTT relation. The answer is yes. The calculation
will appear elsewhere.\\

We would like to thank Prof. Y.S. Wu and Prof. F.D.M. Haldane for
their enlighten discussions during the 7-th Nankai Workshop.
This work was supported in part by the National
Natural Science Foundation of China.


\newpage
\begin{thebibliography}{99}
\bibitem{r1}V.G. Drinfeld, {\em Sov. Math. Dokl.} {\bf 32}, 254-258(1985).
\bibitem{r2} V.G. Drinfeld, Quantum Group, {\em Proc. of the ICM-86},
(Berkeley) Vol.1 New York. Academic press 1986, pp789-820.
\bibitem{r3} V.G. Drinfeld, {\em Sov. Math. Dokl.} {\em 36}, 212-216(1988).
\bibitem{r4}J. Ding and I.B. Frenkel, {\em Commun. Math. Phys.}
{\bf 156}, 277-300(1993).
\bibitem{r5}E.K. Sklyanian, Quantum inverse scattering method, selected
topics from ``Quantum Group and Quantum Integrable Systems'', World
Scientific, Singapore(1991), pp.63-88;\\
L.D. Faddeev, N.Yu Reshetikhin, L.A. Takhtajan, {\em Algebra and Analysis},
{\em 1.1}, 118-206(1989)(in Russian);\\
N.Yu. Reshetikhin, M.A. Semenov-Tian-Shansky, {\em Lett. Math. Phys.}
{\bf 19}, 133-142(1990).
\bibitem{r6}M. Jimbo, ``Introduction to the Yang-Baxter equation'',
Advanced series in mathematical physics, Vol.9, World Scientific(Singapore)
1989, pp.111-134. See the references therein.
\bibitem{r7}P.P. Kulish and E.K. Sklyanin, {\em Lecture Notes in Physics}
{\bf V.151}, 61-119(1982).
\bibitem{r8} C.N. Yang, {\em Phys. Rev. Lett.} {\bf 19}, 1312(1967);
   {\em Phys. Rev.} {\bf 168}, 1920(1968);\\
  R.J. Baxter, {\em Exactly solved models in statistical
   mechanics}, (Academic, London, 1982).
\bibitem{r11} F.D.M. Haldane, {\em Phys. Rev. Lett.} {\bf 60}, 635(1988).
\bibitem{r12} B.S. Shastry, {\em Phys. Rev. Lett.} {\bf 60}, 639(1988).
\bibitem{r13} A. Polychronakos, {\em Phys. Rev. Lett.} {\bf 69}, 703(1992).
\bibitem{r14} J.A. Minahan and A. Polychronakos, {\em Phys. Lett.}
            {\bf B302}, 265(1993).
\bibitem{r15} B. Sutherland and B.S. Shastry, {\em Phys. Rev. Lett.}
  {\bf 71}, 5(1993).
\bibitem{r16} Z.N.C. Ha and F.D.M. Haldane, {\em Phys. Rev.}
   {\bf B46}, 9359(1992).
\bibitem{r17} K. Hikami and M. Wadati, {\em J. Phys. Soc. Japan}
  {\bf 62}, 469(1993).
\bibitem{r18} D. Bernard, M. Gaudin, F.D.M. Haldane and V. Pasquier,
  {\em J. Phys.} {\bf A26}, 5219 (1993).
\bibitem{rc}F. Calogero, O.Ragnisco and C. Marchioro, {\em Lett. Nuovo Cimento}
{\bf 13}, 383(1975).
\bibitem{r19} B. Sutherland and R.A. R\"{o}mer, {\em Phys. Rev. Lett.}
{\bf 71}, 2789(1993).
\bibitem{r20} M.L. Yan, Z. Chen, ``Rational solution of Yang-Baxter
   equation and 1D many-body problem with generalized Delta-function'',
   1989, preprint.
\bibitem{r21} I.V. Komarov, {\em Theor. Mat. Fiz.} {\bf 50}, 402(1982).\\
I.V. Komarov and V.V. Zalipaev, {\em J. Phys.} {\bf A17}, 31(1984).
\bibitem{r22} E.K. Sklyanian, {\em Zapiski Nauchnykh Seminarov Leningradskogo
Otdeleniya Mat. Instit. im. V. A. Steklova AN SSSR} {\bf 133}, 236(1984);\\
Nankai Lectures on Mathematical Physics: {\em Quantum Group and Quantum
Integrable Systems}, (World Scientific, 1991, Singapore), pp63-68.
\bibitem{r23} M.-L. Ge and K. Xue,``Truncated $SL_q(2)$ Algebra:
Trigonometric Goryachev-Chaplygin Gyrostat'',  preprint, 1994.
\end{thebibliography}
\end{document}